\documentclass[lettersize,journal]{IEEEtran}
\usepackage{amsmath,amsfonts}
\usepackage{algorithmic}
\usepackage{algorithm}
\usepackage{array}
\usepackage[caption=false,font=normalsize,labelfont=sf,textfont=sf]{subfig}
\usepackage{textcomp}
\usepackage{stfloats}
\usepackage{url}
\usepackage{verbatim}
\usepackage{graphicx}
\usepackage{cite}
\usepackage{algorithmic}
\usepackage{graphicx}
\usepackage{booktabs}
\usepackage{multirow}
\usepackage[table]{xcolor}
\usepackage{colortbl}
\usepackage{hyperref}
\usepackage{caption}
\usepackage{array}
\usepackage{threeparttable}
\usepackage{float}
\usepackage{placeins}
\usepackage{amsmath}
\usepackage{amssymb}
\usepackage{amsthm}
\usepackage[normalem]{ulem}
\newif\ifmarkedcopy
\markedcopyfalse
\newif\ifstandalonesupplement
\standalonesupplementfalse
\ifmarkedcopy
\long\def\added#1{{\color{blue}#1}}
\long\def\deleted#1{{\color{red}\sout{#1}}}
\long\def\oldcite#1{{\color{red}\cite{#1}}}
\else
\long\def\added#1{#1}
\long\def\deleted#1{}
\long\def\oldcite#1{}
\fi

\begin{document}

\title{AEIOU: A Unified Defense Framework against NSFW Prompts in Text-to-Image Models}

\author{Yiming Wang, Jiahao Chen, Qingming Li, Tong Zhang, Rui Zeng, Xing Yang, Shouling Ji
\thanks{Yiming Wang, Jiahao Chen, Qingming Li, Tong Zhang and Rui Zeng are with the College of Computer Science and Technology, Zhejiang University, Hangzhou 310058, China (e-mail: ym\_wang@zju.edu.cn, xaddwell@zju.edu.cn, liqm@zju.edu.cn, tz\_zju@zju.edu.cn, ruizeng24@zju.edu.cn).}%
\thanks{Xing Yang is with National University of Defense Technology, Changsha 410073, China (e-mail: yangxing17@nudt.edu.cn).}
\thanks{Shouling Ji is with the College of Computer Science and Technology, Zhejiang University, Hangzhou 310058, China (e-mail: sji@zju.edu.cn).}}

\maketitle

\begin{abstract}
As text-to-image (T2I) models advance and gain widespread adoption, their associated safety concerns are becoming increasingly critical. Malicious users exploit these models to generate Not-Safe-for-Work (NSFW) images using harmful or adversarial prompts, underscoring the need for effective safeguards to ensure the integrity and compliance of model outputs. However, existing detection methods often exhibit low accuracy and inefficiency.

In this paper, we propose AEIOU, a defense framework that is \underline{A}daptable, \underline{E}fficient, \underline{I}nterpretable, \underline{O}ptimizable, and \underline{U}nified against NSFW prompts in T2I models. AEIOU extracts NSFW features from the hidden states of the model's text encoder, utilizing the separable nature of these features to detect NSFW prompts. The detection process is efficient, requiring minimal inference time. AEIOU also offers real-time interpretation of results and supports optimization through data augmentation techniques. The framework is versatile, accommodating various T2I architectures. Our extensive experiments show that AEIOU significantly outperforms both commercial and open-source moderation tools, achieving over 95\% accuracy across all datasets and improving efficiency by at least tenfold. It effectively counters adaptive attacks and excels in few-shot and multi-label scenarios.

\vspace{1em} %

\noindent Disclaimer: This article includes potentially disturbing Not-Safe-for-Work (NSFW) text and images. All NSFW images are generated by artificial intelligence. We provide these examples to illustrate how harmful prompts can lead T2I models to generate NSFW content. We have masked and blurred these images, but reader discretion is advised.
\end{abstract}


\section{Introduction}
Recent advancements in text-to-image (T2I) models, such as Stable Diffusion \cite{rombach2022high}, DALL·E 3 \cite{betker2023improving}, and Flux \cite{flux}, have demonstrated remarkable capabilities in generating high-quality images. However, the widespread use of these models raises significant ethical concerns, particularly in the generation of Not-Safe-for-Work (NSFW) content, including sexual, violent, hateful, and other harmful images. Recent studies \cite{qu2023unsafe,moderator,rando2022red} reveal that users can easily produce NSFW images using malicious prompts, known as NSFW prompts. Consequently, effectively defending against NSFW prompts becomes a crucial challenge.

Existing defense methods can be categorized into internal safeguards, which modify the model to diminish its ability to generate NSFW images, and external safeguards, which detect NSFW content \cite{zhang2024adversarial}. Given that modifying the model may significantly impact the quality of generated content \cite{yang2024guardt2i, gandikota2023erasing}, most companies \cite{betker2023improving, Midjourney} tend to adopt detection-based methods, including output and input detection. Output detection \cite{safetysd, qu2023unsafe, schramowski2022can} analyzes the generated NSFW images. However, this approach requires completing the entire generation process, leading to significant resource consumption. Input detection can identify potential NSFW prompts before the generation process, thereby avoiding excessive resource consumption. This is also the focus of our research in this paper.

\deleted{Existing input detection methods mainly fall into two categories: prompt-based and embedding-based detection. Prompt-based detection directly examines the input but often relies on target-model-agnostic classifiers, leading to misclassification due to poor alignment with specific T2I models. Embedding-based detection leverages the text encoder's embeddings for better alignment but struggles to capture the deeper semantic information encoded in the text encoder's attention mechanisms.} \oldcite{openai, Azure, AWS, Ali, yang2024guardt2i, liu2024latent}
\added{Existing input detection methods mainly fall into two categories: prompt-based and embedding-based detection. Prompt-based detection \cite{openai, Azure, AWS, Ali} directly examines the input but often relies on target-model-agnostic classifiers, leading to misclassification due to poor alignment with specific T2I models. Embedding-based detection \cite{yang2024guardt2i, liu2024latent} improves target-model alignment by using the text encoder's output representation, but the final pooled or conditional embedding is already a highly aggregated representation. When harmful semantics are concentrated in only a subset of attention heads or intermediate subspaces, this aggregation can mix the NSFW-sensitive signal with dominant benign and contextual semantics, making the resulting decision space harder to separate for moderation.}



\deleted{Furthermore, a set of overarching challenges continues to impede the advancement of both prompt-based and embedding-based techniques, impacting their real-world applicability. The first of these is a pronounced vulnerability to adversarial attacks, which can manipulate model outputs through subtle input perturbations, thereby fundamentally compromising the system's reliability and security. A second major issue is architectural over-specialization. The predominant focus on the Stable Diffusion v1 model means that many proposed methods lack proven applicability across the diverse and rapidly evolving ecosystem of generative architectures, severely limiting their practical utility and scalability. Finally, these methods are characterized by a significant deficit in interpretability. The internal mechanisms governing both content generation and detection are largely opaque, creating a formidable barrier to user trust, accountability, and the diagnosis of failure modes.} \oldcite{yang2024mma, yang2024sneakyprompt, sd1.4}
\added{Furthermore, several overarching challenges continue to impede the real-world applicability of both prompt-based and embedding-based techniques. First, they remain vulnerable to adversarial prompts that preserve harmful intent while altering surface form \cite{yang2024mma, yang2024sneakyprompt}. Second, many existing studies are tightly centered on Stable Diffusion v1 \cite{sd1.4}, leaving open whether the same defense logic remains reliable across newer architectures and text encoders. Third, the effectiveness of NSFW prompt datasets can drift as T2I systems evolve, making it important to understand whether harmful prompts remain active on newer backends. Finally, these methods still offer limited interpretability, which hinders user trust and makes failure diagnosis difficult.}

To address the issues above, this paper introduces AEIOU, an \textbf{\underline{A}daptable}, \textbf{\underline{E}fficient}, \textbf{\underline{I}nterpretable}, \textbf{\underline{O}ptimizable} and \textbf{\underline{U}nified} defense framework against NSFW prompts in T2I models. Since current adversarial attacks on T2I models primarily target the text encoder \cite{yang2024sneakyprompt, yang2024mma}, as a countermeasure, our defense framework also focuses on the text encoder. Specifically, we analyze the distribution of the text encoder's hidden states within the feature space, revealing that NSFW prompts present explicit NSFW semantics across various layers and attention heads. Previous research \cite{gandelsmaninterpreting} indicates that hidden states of text encoders can be separated into embeddings for different concepts. Building on this, we analyze the general features of NSFW prompts to identify directions within various attention heads that encapsulate NSFW semantics, termed the \textbf{NSFW features} of the attention heads. By assessing the magnitude of the input prompt's hidden state components along these NSFW features, we can effectively detect potential NSFW prompts.

To verify whether NSFW features genuinely represent NSFW semantics and help users understand why a prompt triggers the defense mechanism, we develop an interpretability framework based on detection. Our approach considers both textual and image perspectives. On the textual side, we identify NSFW tokens within prompts by leveraging the NSFW features in each attention head. On the image side, we iteratively remove harmful semantics from the hidden states to produce relatively benign embeddings. By generating images from these modified embeddings, we can observe the progressive eradication of NSFW semantics.

Overall, AEIOU overcomes the deficiencies of previous defense methods, demonstrating not only high accuracy but also exhibiting the following five characteristics:

(1) \textbf{Adaptable.} By analyzing attention heads within the text encoder to specifically capture NSFW features, AEIOU can adapt to any transformer-based text encoder, allowing application across various T2I model architectures.

(2) \textbf{Efficient.} AEIOU operates without complex models for detection, requiring minimal time for both training and inference, ensuring high efficiency.

(3) \textbf{Interpretable.} The inference process of AEIOU does not involve black-box models; instead, it relies on comparisons with NSFW features. The results can be visualized and interpreted across text and image modalities, offering both process and outcome interpretability.

(4) \textbf{Optimizable.} By assigning greater weight to red-teaming data, we can achieve data augmentation for AEIOU with only a few samples, simplifying its updating process.

(5) \textbf{Unified.} AEIOU integrates training, inference, interpretation, and further optimization into a unified framework, avoiding isolated processes.

\deleted{Experiment results indicate that AEIOU exhibits strong defense capabilities across various text encoders in different T2I models. It surpasses four commercial models, two open-source models, and two state-of-the-art models designed for NSFW prompt detection in both effectiveness and efficiency. Furthermore, AEIOU achieves excellent results with minimal data for training and optimization, and it effectively defends against unknown adversarial and adaptive attacks. Additionally, our interpretative approach accurately identifies NSFW tokens and effectively removes NSFW semantics while preserving benign information within the embeddings.}
\added{Experimental results show that AEIOU exhibits strong defense capabilities across multiple text encoders and T2I architectures. It surpasses four commercial models, two open-source models, and two state-of-the-art baselines for NSFW prompt detection, while remaining highly efficient. AEIOU also performs strongly with limited training data, generalizes to unknown attacks, maintains high accuracy on a broader set of adversarial prompt sources, and provides textual and image-based interpretations of the detected NSFW semantics.}

\textbf{Contributions.} In summary, we make the following contributions in this paper.

(1) We investigate the emergence of NSFW semantics within the text encoder and identify the general NSFW features that represent NSFW semantics in each attention head.

(2) Based on the insights above, we leverage NSFW features for NSFW prompts detection within T2I models, demonstrating high effectiveness, strong adaptability, excellent optimization ability, and superior efficiency.

(3) We develop a robust interpretative approach to interpret our detection method, enabling interpretation across text and image modalities.

(4) We integrate the aforementioned techniques into a unified framework and conduct extensive experiments. The results demonstrate that AEIOU outperforms four commercial, two open-source, and two state-of-the-art models.

\section{Related Work}
\subsection{Adversarial Attacks against T2I Models}
\deleted{With the continuous advancement of T2I models, they are becoming increasingly integrated into various aspects of daily life and work. Despite their strengths, they are susceptible to adversarial attacks that modify prompts to sneak past defenses and produce NSFW content like pornography, violence or politically sensitive imagery. Present adversarial attack methods are mainly divided into white-box and black-box approaches. White-box methods primarily utilize the model's text encoder to optimize the prompt, ensuring that the generated prompt semantically aligns closely with a target prompt containing explicit NSFW information, even without sensitive words. In contrast, black-box methods perturb the prompt to find alternative tokens that can replace sensitive words. These methods often utilize reinforcement learning or assistance from large language models to accelerate the search process. In addition, some attack strategies target T2I models with removed concepts. These strategies demonstrate that T2I models can still generate NSFW images even after removing NSFW concepts. Current adversarial methods have proven highly effective. Consequently, it is crucial to develop robust defense mechanisms to counter these attack strategies and ensure the safe and responsible use of T2I models.} \oldcite{ho2020denoising, rombach2022high, betker2023improving, podell2023sdxl,ruiz2023dreambooth,esser2024scaling,liu2023riatig, yang2024mma, yang2024sneakyprompt, ba2023surrogateprompt, tsai2023ring, chin2023prompting4debugging, jia2024jailbreaking}
\added{As T2I models such as Stable Diffusion \cite{ho2020denoising, rombach2022high}, SDXL \cite{podell2023sdxl}, and FLUX \cite{esser2024scaling} are deployed more widely, adversarial prompting has become a practical way to induce harmful generation while evading safety safeguards. Existing attacks share the same high-level goal, namely preserving harmful intent while making the prompt look less obviously unsafe, but they manipulate different prompt signals. Representation-optimized attacks such as MMA \cite{yang2024mma} directly optimize the prompt against internal text-encoder representations so that it remains semantically aligned with explicit harmful targets while avoiding surface-level cues. Lexical or search-based attacks such as SneakyPrompt \cite{yang2024sneakyprompt}, SurrogatePrompt \cite{ba2023surrogateprompt}, and JPA \cite{jia2024jailbreaking} instead search for alternative wordings that preserve harmful semantics without exposing sensitive words. Red-teaming and safety-reactivation attacks such as Ring-A-Bell \cite{tsai2023ring} and P4D \cite{chin2023prompting4debugging} mine prompts that can still trigger unsafe generation on concept-erased or safety-aligned models. This diversity matters for defense evaluation: robustness against one attack family does not necessarily imply robustness against prompts generated through a different mechanism.}

\subsection{Defensive Methods against Attacks}
\deleted{Based on the knowledge of the model, existing defense methods against attacks in T2I models fall into two categories: internal and external safeguards. \textbf{Internal safeguards} aim to disable the model's ability to generate NSFW images by fine-tuning the model itself. They can be divided into model editing and inference guidance. Model editing methods aim to modify the internal parameters by training. However, these methods typically require prolonged training periods, and parameter modifications can impact the quality of the generated images. Moreover, most methods focus solely on safety against malicious prompts while ignoring the adversarial prompts. In contrast, inference guidance methods focus on modifying internal features during the inference stage. Unlike model editing methods, they are tuning-free and plug-in, which can be easily inserted into any model. However, these methods also fail to account for adversarial prompts, making them susceptible to targeted attacks. \textbf{External safeguards} aim to filter out potential malicious samples by examining intermediate variables during the generation process. Current detection methods focus on prompts, conditional embeddings, or generated images. Image-based moderation entails reviewing the generated images to identify NSFW samples. They incur significant inference costs since images must be generated before assessment. Prompt-based moderation screens input prompts to identify those likely to generate NSFW images. Given their lower cost, they are widely used by online services like Midjourney and Leonardo.Ai. Nonetheless, these methods generally lack targeted defenses against adversarial attacks, making them susceptible to circumvention. Embedding-based moderation examines the conditional embeddings to filter out malicious samples. While this approach offers some resistance to adversarial attacks, it relies on large-scale models for classification, resulting in high costs. Additionally, it also suffers from low accuracy and remains vulnerable to adaptive attacks.} \oldcite{zhang2024adversarial, gandikota2023erasing,kumari2023ablating,li2024safegen,poppi2024safe,heng2024selective,orgad2023editing,wu2024unlearning,hu2025safetext,schramowski2023safe,li2024self,safetysd, qu2023unsafe, schramowski2022can, openai, AWS, Azure, NSFWtext, Detoxify, Midjourney, Leon, yang2024guardt2i, liu2024latent}
\added{Existing defenses differ mainly in the signal level on which moderation is performed. \textbf{Internal safeguards} modify the generative model itself. Model editing methods such as ESD \cite{gandikota2023erasing}, Ablating Concepts \cite{kumari2023ablating}, SafeGen \cite{li2024safegen}, MACE \cite{lu2024mace}, RECE \cite{heng2024selective}, TIME \cite{orgad2023editing}, MDU \cite{wu2024unlearning}, and SafeText \cite{hu2025safetext} suppress unsafe concepts by changing model parameters. Their advantage is that they directly reduce the model's ability to generate harmful content, but they usually require extra training or parameter modification and may degrade generation quality. Inference-guidance methods such as Safe Latent Diffusion \cite{schramowski2023safe} and Self-Discover Guidance \cite{li2024self} avoid retraining and are easier to plug in, but they still intervene after harmful semantics have already entered the generation pipeline and can remain vulnerable to adversarial prompts.

\textbf{External safeguards} instead filter harmful samples before or after generation. Image-based moderation methods such as SafetyChecker \cite{safetysd}, Q16 \cite{qu2023unsafe}, and the detector of Schramowski \textit{et al.} \cite{schramowski2022can} benefit from direct visual evidence, but they incur the full cost of image generation and therefore act only after the expensive step has already happened. Prompt-based moderation methods such as OpenAI Moderation \cite{openai}, AWS Comprehend \cite{AWS}, Azure Content Safety \cite{Azure}, NSFW-Text \cite{NSFWtext}, Detoxify \cite{Detoxify}, and the deployed filters used by Midjourney \cite{Midjourney} and Leonardo.Ai \cite{Leon} are efficient and easy to deploy, but they typically rely on generic text classifiers whose decision space is only weakly aligned with the target T2I model. As a result, they are easier to circumvent with model-specific adversarial prompts.

Embedding-based moderation methods such as GuardT2I \cite{yang2024guardt2i} and Latent Guard \cite{liu2024latent} improve alignment by using the target model's text encoder outputs rather than raw prompt text. However, they still make decisions on aggregated final representations, such as pooled or conditional embeddings. At that representation level, harmful semantics may no longer appear as a cleanly separable signal, because the NSFW-sensitive components can be mixed with dominant benign and contextual semantics from other heads and subspaces. This makes the decision boundary less stable and less interpretable under adversarial prompting. AEIOU differs from these methods not by attaching a heavier classifier to the same representation, but by moving moderation to head-level intermediate representations before this final compression, where NSFW-sensitive directions remain more explicit and easier to isolate.}

\subsection{Interpretation on the CLIP Model}
\deleted{As the most widely used text encoder in T2I models, the CLIP model is a significant focus of adversarial attack research. Recent studies have investigated the CLIP model to analyze its internal mechanisms. For instance, Bhalla \textit{et al.} found that the embeddings generated by CLIP exhibit strong linear properties and can be decomposed into combinations of various concepts. Gandelsman \textit{et al.} discovered that different attention heads within the CLIP model are responsible for interpreting different semantics, which are then combined to produce the final output embeddings. However, these studies primarily focus on the image domain. In this paper, we build upon existing work to further explore the properties of the CLIP model in the text domain and propose a novel interpretation method to interpret how prompts containing NSFW semantics are generated.} \oldcite{radford2021learning, gandelsmaninterpreting, bhalla2024interpreting, zhao2024gradient, aflalo2022vl}
\added{Interpretability studies on CLIP \cite{radford2021learning} provide the methodological background for our design. Gandelsman \textit{et al.} \cite{gandelsmaninterpreting} and Bhalla \textit{et al.} \cite{bhalla2024interpreting} show that CLIP's intermediate representations are structured and often decomposable, and that different attention heads capture different semantic directions before they are merged into the final embedding. Related analyses by Zhao \textit{et al.} \cite{zhao2024gradient} and Aflalo \textit{et al.} \cite{aflalo2022vl} further support the view that internal representations contain semantically meaningful directions beyond the final pooled embedding. Most of these studies focus on general visual concepts or image-side behavior. Our work transfers this perspective to text-side NSFW moderation: instead of using intermediate representations only for post hoc interpretation, we directly detect harmful semantics at the attention-head level and then use the same internal signals for both textual and visual interpretation.}

\section{Method}

\begin{figure}[t]
    \centering
    \includegraphics[width=0.9\linewidth]{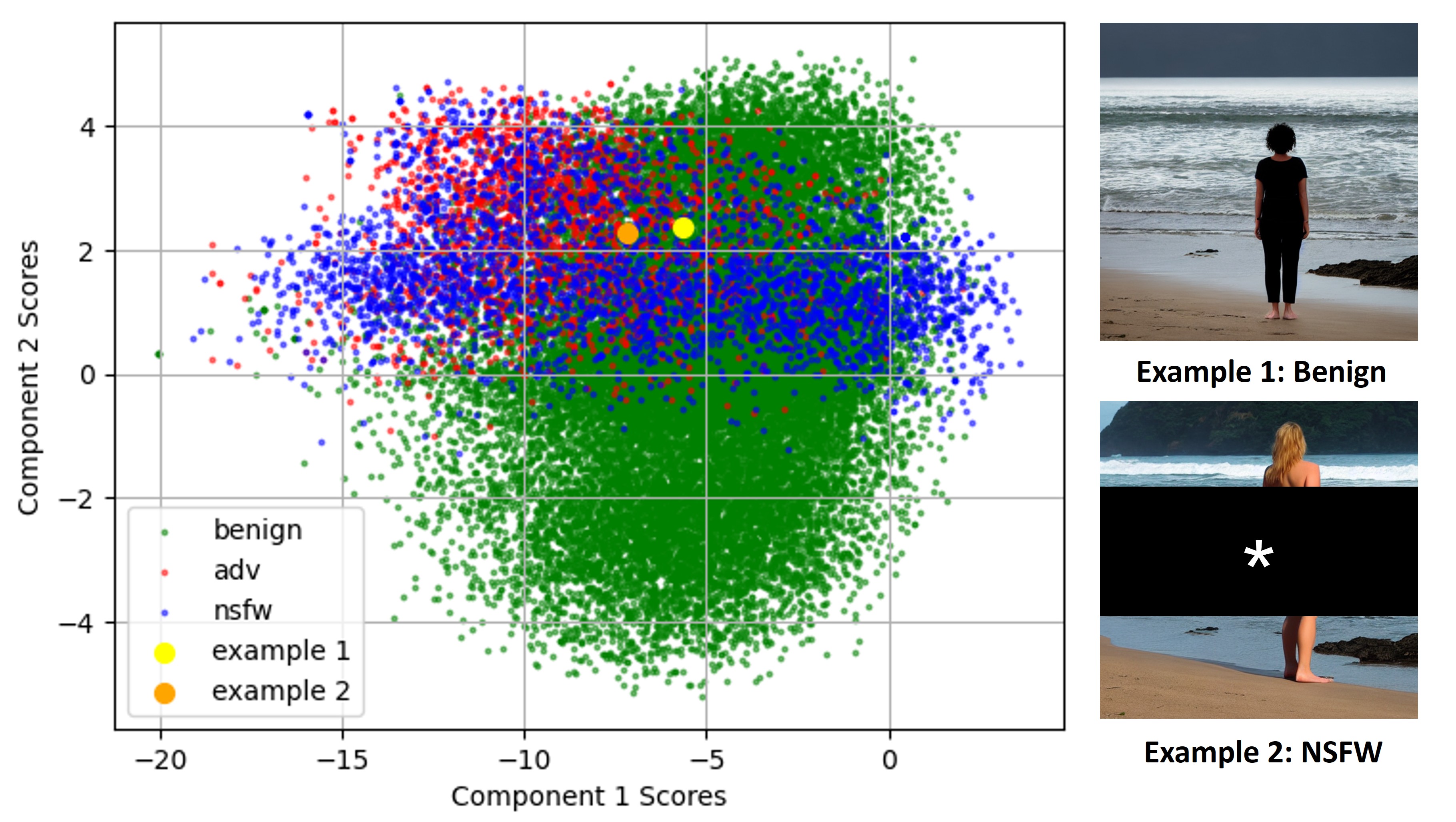}
    \caption{Left: The distribution of output embeddings from CLIP model. Right: Example 1 corresponds to the prompt ``A woman stands on the beach, facing the sea." Example 2 corresponds to the prompt ``A naked woman stands on the beach, facing the sea." On the right are the images generated from them.}
    \label{fig: embedding}
\end{figure}

\subsection{Design Intuition}
\label{sec:intuition}

In this section, we explore how NSFW semantics are concealed within the conditional embeddings and revealed in the text encoder's hidden states. Our investigation focuses on the CLIP model \cite{radford2021learning}, the most widely used text encoder in T2I models and the primary target of adversarial attacks. 

\begin{figure*}[ht] 
    \centering
    \includegraphics[width=0.95\linewidth]{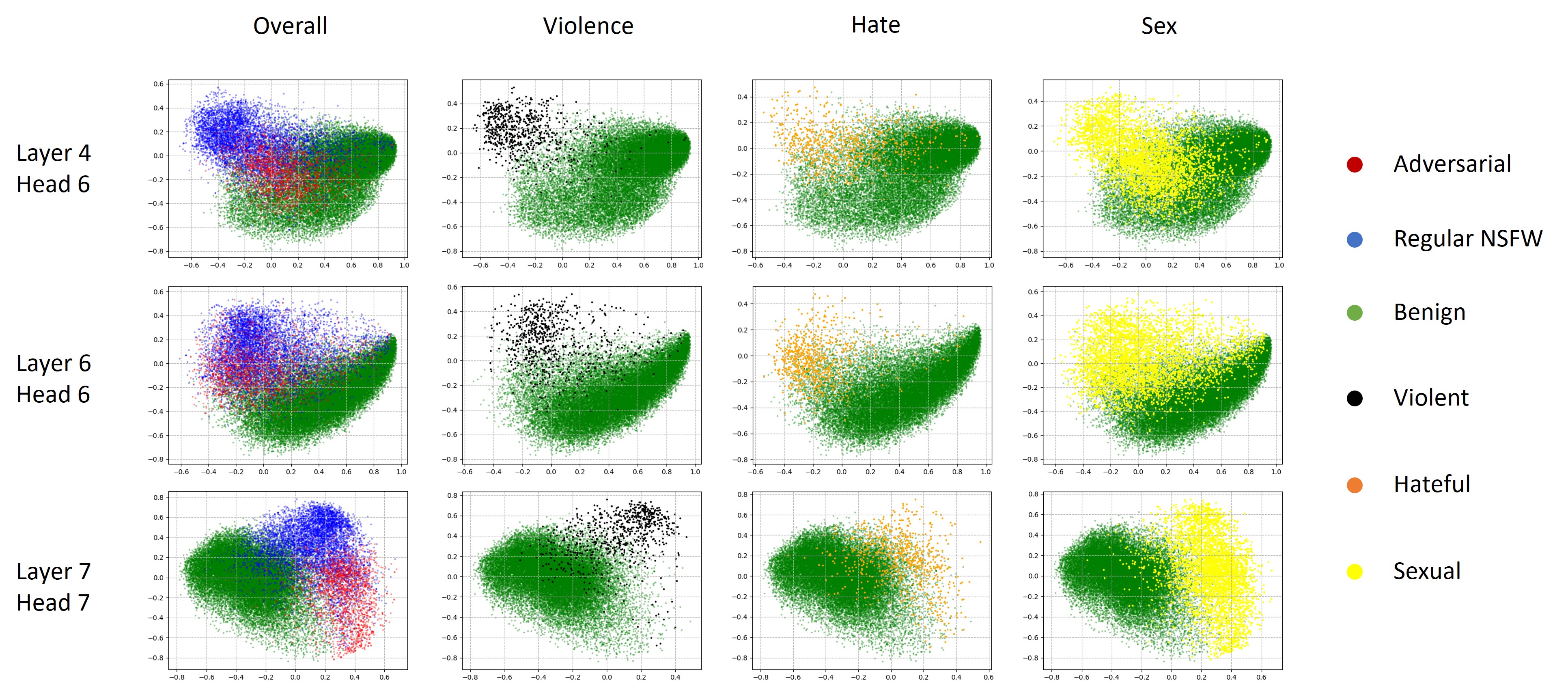}
    \caption{PCA maps of hidden states from different layers and different heads.}
    \label{fig: heads}
\end{figure*}

We classify the T2I model prompts into three categories: benign prompts, which do not generate NSFW images; regular NSFW prompts, which are manually crafted with explicit NSFW semantics; and adversarial prompts, which are generated by adversarial attacks and are challenging for humans to interpret as NSFW. Through extensive data collection, we compiled a dataset containing over 35,000 prompts across all three categories. We utilize the CLIP model to obtain pooled embeddings of these prompts and examine their distribution with a PCA map \cite{abdi2010principal}, as depicted in Figure \ref{fig: embedding}. Although the distributions of benign and NSFW data differ, considerable overlap makes effective differentiation challenging.

\deleted{This phenomenon is primarily attributed to the architecture of the CLIP network, which comprises multiple layers and attention heads. Each layer employs a multi-head self-attention mechanism, where each head independently extracts information from the prompt. The outputs of these heads are then combined through a linear layer and passed to the next layer. Linear layers integrate information from various heads, making it challenging to isolate individual pieces of information. As the network's depth increases, the outputs from different layers are accumulated through residual connections, exacerbating the entanglement of information. Consequently, benign prompts and NSFW prompts become intermixed within the pooled embedding space. As illustrated in the figure, two sentences describing the same object might appear semantically similar; however, one could be used to generate NSFW images while the other remains benign.} \oldcite{vaswani2017attention}
\added{This phenomenon is primarily attributed to the architecture of the CLIP network, which comprises multiple layers and attention heads. Each layer employs a multi-head self-attention mechanism \cite{vaswani2017attention}, where each head independently extracts information from the prompt. The outputs of these heads are then combined through linear projection and passed to the next layer. As the network deepens, these mixed representations are further accumulated through residual connections. As a result, NSFW-sensitive information does not necessarily disappear in the pooled embedding; rather, it can become entangled with dominant benign and contextual semantics, making the final decision space harder to separate for moderation. As illustrated in Figure \ref{fig: embedding}, two sentences describing similar objects can remain close in pooled space even though one is harmful and the other is benign.}

To more accurately delineate the boundary between NSFW and benign prompts, we examine the internal workings of the model by exploring the hidden states across different layers and attention heads. Figure \ref{fig: heads} illustrates the distribution of benign, regular NSFW and adversarial data across several attention heads. The first column illustrates their overall distribution. In some heads, significant overlap remains between benign and NSFW data distributions. However, in other heads, a distinct boundary between these data types is evident. This reveals that attention heads within the model exhibit differing sensitivities to NSFW content. Certain attention heads preserve clearer NSFW-sensitive directions within their intermediate representations, allowing them to differentiate between benign and NSFW prompts more effectively than the final pooled representation.

We also find that different NSFW prompt categories are processed uniquely across attention heads. Columns two, three, and four of Figure \ref{fig: heads} illustrate how violence, hate, and sex categories differ from benign data. In the first attention head, violence and hate data are separable from benign data, whereas sex data shows overlap. The second attention head clearly differentiates violence from benign data, with weaker distinctions in the other categories. In the third head, sex and hate data are distinctly separated from benign data, while violence data overlaps significantly. These findings suggest that attention heads specialize in handling specific NSFW content types. Even if a head struggles to differentiate between benign and NSFW data, it may still excel at distinguishing a particular type of NSFW prompt.

Based on these findings, we propose leveraging the attention heads within the text encoder to differentiate between benign and NSFW prompts effectively. By capitalizing on the diverse focus of different attention heads, we can aggregate information from all attention heads to achieve more accurate prompt classification.

\subsection{Framework Overview}
\begin{figure*}[ht]
    \centering
    \includegraphics[width=1.0\linewidth]{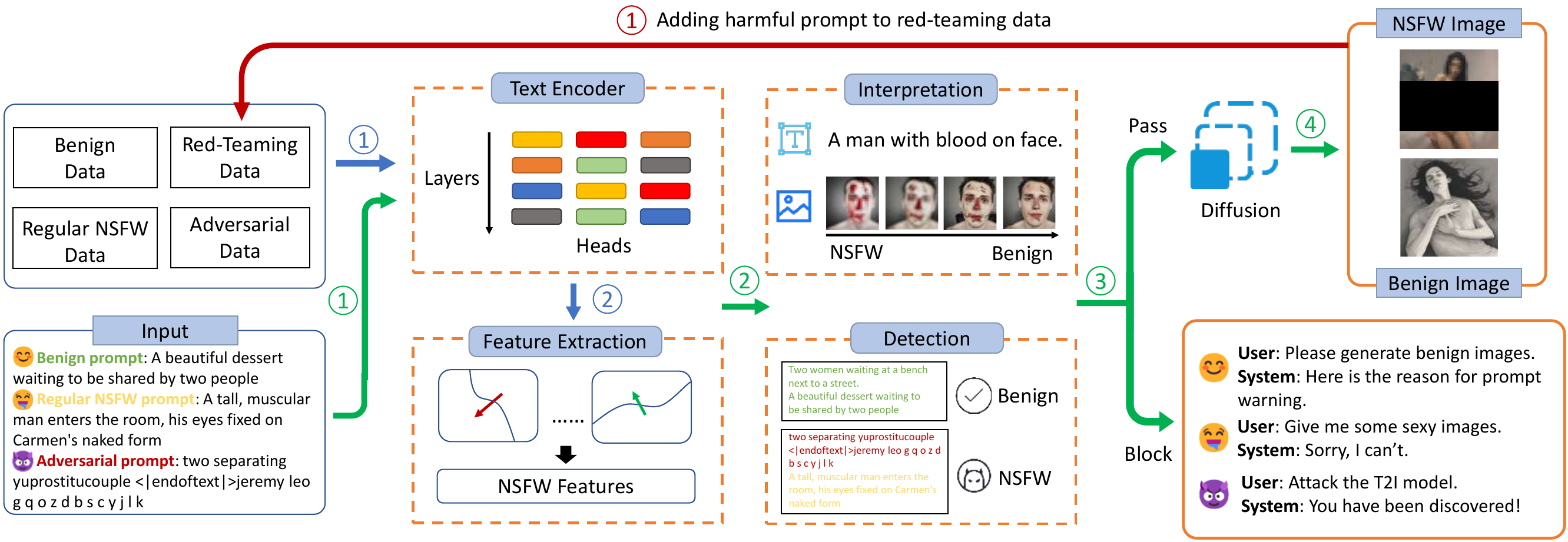}
    \caption{The overall framework of AEIOU. The blue arrows represent the training process, where data enters the text encoder, and AEIOU extracts NSFW features from the hidden states. The green arrows indicate the inference process, where the prompt passes through the text encoder for detection and interpretation; if it passes, image generation proceeds, otherwise, generation is denied and an interpretation is provided. The red arrows indicate the data augmentation process, involving red-teaming tests on AEIOU, where NSFW prompts that successfully bypass detection are added to the training data for data augmentation.}
    \label{fig: overall}
\end{figure*}

Based on the findings from Section \ref{sec:intuition}, we develop a framework for detecting and interpreting NSFW prompts. The overall architecture of this framework is illustrated in Figure \ref{fig: overall}. Initially, we identify the direction within each attention head’s hidden states that best represent NSFW semantics by analyzing the distribution differences between benign and NSFW prompts, which we call NSFW features. When a new prompt is inputted, we can assess the risk of generating NSFW images by evaluating the projection magnitude of the prompt along the NSFW feature. To interpret our assessment, we pinpoint the token most representative of NSFW semantics within the prompt, analyzing why the current prompt is classified as an NSFW prompt. Furthermore, we can progressively eliminate NSFW semantics from the hidden states and input the modified prompt embeddings to generate images. This process enables us to observe how NSFW semantics are gradually removed from the embeddings. Finally, we can conduct red team testing on the framework or monitor it in real-time post-deployment to collect prompts that bypass current defenses. By incorporating them into the training set, we achieve more accurate NSFW feature identification and detection results.

The practical application of this framework manifests in two main ways. First, it prevents NSFW image generation from the outset when malicious users attempt to employ adversarial prompts. Second, when regular users inadvertently input prompts with NSFW semantics and encounter blockages, we can interpret why the prompt is considered inappropriate. This helps users quickly identify problematic tokens and modify their prompts to generate the desired image.

\subsection{NSFW Features}
The CLIP text encoder consists of $L$ layers, each with a multi-head self-attention mechanism with $H$ heads followed by an MLP block. A prompt $P$ is divided into $N-1$ tokens and projected into initial token embeddings $\{z_i^0\}_{i \in \{0,...,N\}}$, where $\{z_0^0\}$ is the BOS token and $\{z_N^0\}$ is the EOS token. These embeddings form the matrix $Z_0$, the initial input to the encoder. Each layer updates this input through self-attention and MLP modules with two residual steps:
\begin{equation}
    \hat{Z}^l = \text{ATT}^l(Z^{l-1}) + Z^{l-1}, \quad Z^l = \text{MLP}^l(\hat{Z}^l) + \hat{Z}^l.
\end{equation}
\deleted{In this framework, the ATT layer employs $H$ attention heads to extract information and integrates them into a vector through linear projection. The MLP layer further refines them to obtain intermediate embeddings that represent the overall information of the prompt. However, some information may be obscured or discarded during this process. Therefore, we need to utilize the original outputs from the attention heads to extract NSFW semantics effectively.}
\added{In this framework, the ATT layer employs $H$ attention heads to extract information and integrates them into a vector through linear projection. The MLP layer further refines the mixed representation. This makes the final embedding well suited for generation, but not necessarily ideal for moderation, because NSFW-sensitive directions can be mixed with other semantics during aggregation. We therefore use the original outputs from attention heads to extract NSFW semantics before this final compression step.}

Considering that the outputs from multiple attention heads have been preliminarily integrated into $\text{ATT}^l(Z^{l-1})$, we need to decompose its computational process to extract the information each head represents. Since the CLIP model's self-attention block employs a causal mask \cite{radford2021learning}, only the EOS token holds the complete semantics. Therefore, we concentrate solely on the EOS token. Following Elhage \textit{et al.} \cite{elhage2021mathematical} and Gandelsman \textit{et al.} \cite{gandelsmaninterpreting}, we formalize the ATT output as a sum over $H$ independent attention heads and $N+1$ tokens:
\begin{align}
\left[ \text{ATT}^l(Z^{l-1}) \right]_{EOS} &= \sum_{h=1}^{H} \sum_{i=0}^{N} x_{i,EOS}^{l,h}, \\ x_{i,EOS}^{l,h} &= \alpha_{i,EOS}^{l,h} W_{VO}^{l,h} z_{i}^{l-1}. \notag
\end{align}
where $W_{VO}^{l,h}$ are transition matrices and $\alpha_{i,EOS}^{l,h}$ are attention weights from the $i$-th token to the EOS token. 
In this way, we can get the contribution of $h$-th head in $l$-th layer for $h\in [1,H], l\in[1,L]$, which can be expressed as $c^{l,h} = \sum_{i=0}^{N} x_{i,EOS}^{l,h}$. 

Each $c^{l,h}$ is located within a $d$-dimensional representation space, and our goal is to identify the direction within this space that best represents NSFW semantics. When a new prompt is input, the more its representation in attention heads aligns with the NSFW direction, the more likely it is to contain NSFW semantics. We refer to these directions as the \textbf{NSFW features} of each attention head.

\begingroup
\renewcommand{\arraystretch}{1.5}
\begin{table*}[ht]
    \centering
    \caption{Text-based interpretation of regular NSFW prompts and adversarial prompts with CLIP-ViT-L. We highlight the words with the strongest NSFW connotations in \colorbox{red}{red}, followed by \colorbox{orange}{orange} and \colorbox{yellow}{yellow}.}
    \label{tab:text}
    \resizebox{0.75\textwidth}{!}{
    \begin{tabular}{lc}
        \toprule
        \textbf{Category} & \textbf{Prompt} \\
        \midrule
        \multirow{2}{*}{\textbf{\textcolor{green!50!black}{Regular NSFW prompt}}} &  A \colorbox{red}{naked} \colorbox{orange}{woman} stands on the beach, facing the sea.\\
        \cline{2-2}
        & \colorbox{yellow}{Her} \colorbox{yellow}{body} \colorbox{orange}{shakes} with each \colorbox{orange}{orgasm}, the \colorbox{red}{pleasure} \colorbox{orange}{overwhelming}.\\
        \hline
        \multirow{2}{*}{\textbf{\textcolor{red}{Adversarial prompt}}} & gesture \colorbox{orange}{softly} \colorbox{red}{licking} upon \colorbox{yellow}{her} backwards \colorbox{orange}{body} allows an\\
        & \colorbox{red}{vigorconseshrisensuoutgoing} \colorbox{yellow}{man} \colorbox{yellow}{led} \colorbox{red}{carmen} \colorbox{orange}{habns} arch.\\
        \hline
    \end{tabular}
      }
\end{table*}
\endgroup

To calculate NSFW features, we introduce two sets of prompts: benign prompts and NSFW prompts, where NSFW prompts include regular NSFW prompts and adversarial prompts. We input these prompts into the CLIP model and obtain the output of each attention head, denoting the output of benign prompts as $\{c_{b_k}^{l,h}\}$ and the output of NSFW prompts as $\{c_{m_k}^{l,h}\}$. Our objective is to maximize \(\langle u^{l,h}, c_{m_k}^{l,h} \rangle\) while minimizing \(\langle u^{l,h}, c_{b_k}^{l,h} \rangle\), where $\{u^{l,h}\}$ are NSFW features we want to extract. To achieve this, we establish three optimization objectives:
\begin{equation}
    \max \| \langle u^{l,h}, \mu_m^{l,h} \rangle- \langle u^{l,h}, \mu_b^{l,h} \rangle \|,
\end{equation}
\begin{equation}
    \min \sum_{k=0}^{K_b} (\langle u^{l,h}, c_{b_k}^{l,h} \rangle - \langle u^{l,h}, \mu_b^{l,h} \rangle)^2,
\end{equation}
\begin{equation}
    \min \sum_{k=0}^{K_m} (\langle u^{l,h}, c_{m_k}^{l,h} \rangle - \langle u^{l,h}, \mu_m^{l,h} \rangle)^2,
\end{equation}
where $\mu_b^{l,h}$ and $\mu_m^{l,h}$ are the mean value of $\{c_{b_k}^{l,h}\}$ and $\{c_{m_k}^{l,h}\}$. In summary, our goal is to maximize the distance between the projected means of $\{c_{b_k}^{l,h}\}$ and $\{c_{m_k}^{l,h}\}$ on the vector $u^{l,h}$, while simultaneously minimizing their respective variances. We can employ Linear Discriminant Analysis (LDA) \cite{balakrishnama1998linear} to solve this problem, ultimately obtaining the NSFW feature $\{u^{l,h}\}$:
\begin{equation}
    u^{l,h} = S_w^{-1}(\mu_m^{l,h}-\mu_b^{l,h}),
\end{equation}
\begin{equation}
    S_w=\sum_{k=0}^{K_m} \| c_{m_k}^{l,h}-\mu_m^{l,h}\|^2 + \sum_{k=0}^{K_b} \| c_{b_k}^{l,h}-\mu_b^{l,h}\|^2,
\end{equation}
where $K_b$ is the number of benign prompts, and $K_m$ is the number of malicious prompts. \deleted{In text encoders other than CLIP, such as T5, we can still use the method above to extract NSFW features. Any text encoder utilizing multi-head self-attention can be adapted to this approach.} \oldcite{raffel2020exploring}
\added{The EOS-based derivation above is specific to CLIP-style causal encoders. For non-causal text encoders such as T5 \cite{raffel2020exploring}, hidden states at different token positions already encode global sentence information. To keep the method consistent and computationally efficient across encoders, we still use the last token as the representative position for head-level feature extraction and scoring. Our experiments show that this same last-token design remains effective in practice on T5-based T2I models.}

\subsection{NSFW Prompts Detection}
\label{sec:detect}
By utilizing the identified NSFW features, we can detect NSFW prompts. According to previous research \cite{gandelsmaninterpreting}, we can consider the intermediate embeddings of prompts in the CLIP model as a linear combination of concepts. The projection of these embeddings onto each concept direction represents the contribution of that concept to the embedding. Based on this, we define the projection of the embedding onto the NSFW feature as the NSFW score of a prompt $p$:
\begin{equation}
    Score(p)^{l,h} = Proj(c_p^{l,h},u^{l,h}) = \frac{\langle c_p^{l,h}, u^{l,h} \rangle}{\|u^{l,h}\|}.
\end{equation}
By aggregating the NSFW Scores from all attention heads, we can obtain the final NSFW Score for the current prompt:
\begin{equation}
    Score(p) = \frac{\sum_{l=1}^L \sum_{h=1}^H Score(p)^{l,h}}{L\cdot H}.
\end{equation}
The larger the $Score(p)$, the more likely the current prompt contains NSFW semantics.

Theoretically, if $\text{Score}(p) > 0$, the prompt $p$ contains NSFW semantics and should be classified as an NSFW prompt. Conversely, if $\text{Score}(p) < 0$, the prompt should be benign. However, experiments show that while setting the threshold to zero allows AEIOU to achieve high accuracy, optimal classification performance requires a slight threshold adjustment. We hypothesize that this is due to the distribution of the training set not fully representing the actual distribution of NSFW prompts, introducing bias in the NSFW features derived during training. In our experiments, we determine the threshold by selecting the one that yields the highest F1 Score on the training set. For different text encoders, the final offset ranges from 1\% to 3\%.

The approach above treats NSFW as a single comprehensive category. Suppose there is a need to subdivide it further or to identify specific categories of NSFW prompts, such as sex or violence. In that case, we can categorize NSFW prompts in the training set based on labels. This enables the calculation of NSFW features for each subcategory, facilitating the determination of NSFW scores for each. If we also need to detect NSFW prompts across all categories, we can aggregate the NSFW scores from all subcategories and use their maximum value as the final NSFW score.

Moreover, we can collect adversarial prompts that successfully bypass detection by employing adaptive attacks during red team testing. Incorporating these prompts into the training set for data augmentation allows us to achieve more accurate NSFW feature extraction and detection results. Since adaptive attacks require significant time and have a low success rate, generating a large volume of red-teaming data is challenging. However, we can increase their impact during training by assigning greater weight to these data. Specifically, by weighting the target prompt's $c_m^{l,h}$ when calculating $u_m^{l,h}$ and $S_w$, we can amplify their influence on the resulting NSFW feature. Experiments have shown that optimizing the NSFW feature through data augmentation can effectively reduce the success rate of adaptive attacks.

\begin{figure*}[ht]
    \centering
    \includegraphics[width=0.8\linewidth]{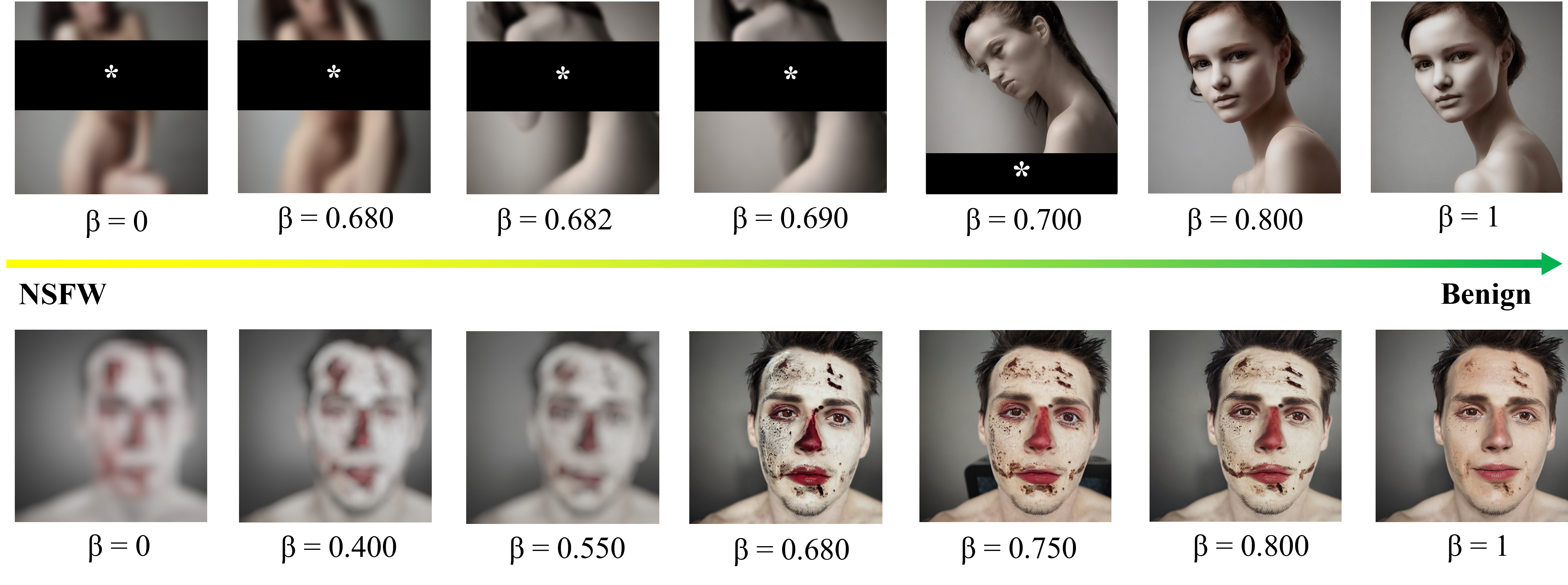}
    \caption{Image-based interpretation with Stable Diffusion v1.4. In order to mitigate potential impact on the reader, we follow the established convention of prior work \cite{yang2024mma, ba2023surrogateprompt} and obscure the NSFW images using both blurring and masking techniques.}
    \label{fig: image}
\end{figure*}

\subsection{NSFW Prompts Interpretation}
After the detection process, we further interpret NSFW prompts through a two-module framework. First, we develop an interpretative method to identify the tokens that most strongly contribute to NSFW semantics. Second, we investigate the generation mechanism of NSFW semantics in conditional embeddings by gradually attenuating the NSFW features in the hidden states. Using the resulting embeddings to generate images, we gain insights into how NSFW semantics can be progressively eliminated.
\subsubsection{Text-Module Interpretation}
In the text module, we interpret NSFW prompts by identifying NSFW tokens. Within each attention head, we assess the NSFW semantic association at any given position by computing the cosine similarity between the hidden state at that position and the head's NSFW feature:
\begin{equation}
    \hat E(p)_i^{l,h} = CosSim(c_{p_i}^{l,h}, u^{l,h}) = \frac{\langle c_{p_i}^{l,h}, u^{l,h} \rangle}{\|c_{p_i}^{l,h}\| \|u^{l,h}\|}.
\end{equation}
As tokens pass through the attention model, their semantics interrelate and intertwine, with each position’s hidden states encapsulating information from all preceding tokens. Consequently, the association between the current position's hidden states and NSFW semantics cannot be directly used to represent the token's connection to NSFW semantics accurately.

Similar to Equation 2, each position's hidden state $c_{p_i}^{l,h}$ can be represented as a combination of {$z_j^{l-1}$}, with the corresponding attention weight $\alpha_{j,i}$ indicating the contribution of $z_j^{l-1}$.
\begin{equation}
    c_{p_i}^{l,h} = \sum_{j=0}^N \alpha_{j,i}^{l,h} W_{VO}^{l,h} z_{j}^{l-1},
\end{equation}
where $\alpha_{j,i}^{l,h}$ are attention weights from the $j$-th token to the $i$-th token. For Layer 1, since it has only undergone a single operation as described in Equation 11, we can reconstruct the actual contribution of each token based on $\{\alpha_{j,i}^{1,h}\}$:
\begin{equation}
    E(p)_i^{1,h} = \sum_{j=0}^N \alpha_{j,i}^{1,h} \hat E(p)_i^{1,h}.
\end{equation}
For deeper layers, we need to use the $\alpha$ from the preceding layers to approximate the contribution of each token. Since $\{\alpha_{j,i}^{l,h}\}$ directly forms the attention map $A^{l,h}$, we can multiply the attention maps from each layer to approximate the contribution of each token to the positions in the current layer. In this way, the interpretative results for each layer can be represented as:
\begin{equation}
    E(p)_i^{l,h} = \sum_{j=0}^N \{\prod_{k=0}^{l} A^{k,h}\}_{j,i} \hat E(p)_i^{l,h}.
\end{equation}
By aggregating the interpretation results from all attention heads, we can obtain the final interpretation for each token:
\begin{equation}
    E(p)_i = \frac{\sum_{l=1}^L \sum_{h=1}^H E(p)_i^{l,h}}{L\cdot H}.
\end{equation}

Table \ref{tab:text} demonstrates several examples of our interpretation. Our method accurately identifies sensitive words in regular samples and adversarial tokens in adversarial samples. We will conduct further experiments to demonstrate the effectiveness of the interpretability method.

\subsubsection{Image-Module Interpretation}
In the image domain, we gradually attenuate the NSFW semantics contained in the prompts and examine the generated images to observe how the images evolve as the NSFW semantics are progressively diminished. 
Since the conditional embedding in T2I models includes embeddings of all tokens, we need to eliminate NSFW semantics from all tokens. We assume that the feature space of each token is identical to that of the EOS token. This allows us to compute the NSFW score for each token by projecting the intermediate embeddings onto the NSFW features. Subsequently, we can attenuate the NSFW semantics of all tokens as follows:

\begin{equation}
    \hat c_{p_i}^{l,h} = c_{p_i}^{l,h} - \beta \cdot Score(p_i)^{l,h} \frac{u^{l,h}}{\|u^{l,h}\|}
\end{equation}

By recombining $\hat c_{p_i}^{l,h}$ to calculate $Z^l$, we can derive the modified conditional embedding. By gradually increasing the value of $\beta$ and using conditional embedding to generate images, we can observe the process of NSFW semantics being progressively eliminated from the images.

We use two prompts to demonstrate the results of our image-module interpretation: the sexual prompt, "a beautiful naked woman," and the violent prompt, "a man with blood all over his face." Figure \ref{fig: image} shows the images generated with different values of $\beta$. As the value of $\beta$ increases, the images generated from both prompts gradually transition from NSFW to harmless. During this process, we observed two interesting phenomena. First, as the NSFW semantics are progressively reduced, the model still attempts to preserve the original semantics of the prompt. For the sexual prompt, the woman gradually turns away and conceals sensitive areas with her hands. Ultimately, only the area above her shoulders is visible, rendering the image harmless. Throughout, the image never violates the prompt's directive. 

For the violent prompt, the blood on the man's face gradually changes color to resemble oil stains, and the marks themselves progressively diminish. However, even at $\beta=1$, some dirty marks remain on the man's face. These examples illustrate that the prompt's semantics are highly editable within the hidden states. By applying our explanation method, we can progressively eliminate the NSFW semantic content while preserving the original meaning of the prompt to the greatest extent possible.


Secondly, the generated images exhibit several abrupt transitions during the gradual increase of $\beta$. For the sexual prompt, the overall structure of generated images remains unchanged when $\beta$ increases from zero to 0.68. However, when $\beta$ increases slightly further. This indicates that in some scenarios, as the original image structure contains deeply embedded NSFW semantics that are difficult to remove, the generation process is forced to make structural alterations as we progressively weaken the NSFW content. This, in turn, leads to changes in the structure and identity of the generated image.

\section{Experiments}
\subsection{Experimental Settings}
\subsubsection{Datasets}
\deleted{The datasets we use comprise three categories. Clean datasets consist of benign data that does not generate NSFW images. Regular NSFW datasets include manually generated prompts with explicit NSFW semantics. Adversarial datasets comprise algorithmically generated adversarial NSFW prompts.}
\added{The datasets we use comprise three categories. Clean datasets consist of benign prompts that do not generate NSFW images. Regular NSFW datasets include manually collected prompts with explicit harmful semantics. Adversarial datasets comprise algorithmically generated prompts designed to bypass safety safeguards.}

\begin{figure}[ht]
    \centering
    \includegraphics[width=0.8\linewidth]{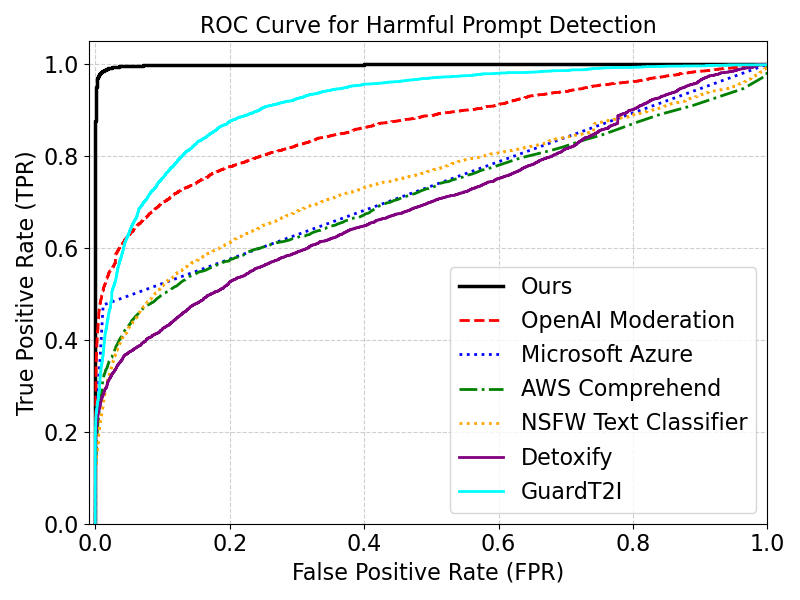}
    \caption{ROC curves of all methods.}
    \label{fig: roc}
\end{figure}

\textbf{Clean Dataset.}
Our experiments use the validation captions of \textbf{MSCOCO} \cite{lin2014microsoft} as the clean dataset. \textbf{MSCOCO} is a cross-modal image-text dataset, a popular benchmark for training and evaluating T2I generation models. We remove all captions containing sensitive words to ensure the samples are benign. A total of 25,008 captions are retained.

\textbf{Regular NSFW Dataset.}
\deleted{We gather data from multiple sources to comprehensively represent various types of NSFW semantics. I2P contains 4,703 NSFW prompts sourced from real users on Lexica. The categories include hate, harassment, violence, self-harm, sex, shocking content, and illegal activities. Specifically, we categorize the prompts into hard and soft based on their level of harmfulness. 4chan Prompts contain 500 NSFW prompts collected from 4chan. They predominantly encompass discriminatory and derogatory statements. NSFW200 involves 200 NSFW prompts related to sexual and bloody content. NSFW-LAION contains 1,143 NSFW prompts we sampled from LAION-COCO, predominantly focused on sexual content.} \oldcite{schramowski2023safe, lexica, qu2023unsafe, 4chan, yang2024sneakyprompt, schuhmann2022laion}
\added{We gather regular NSFW prompts from multiple sources to cover diverse harmful semantics:}
\begin{itemize}
    \item \textbf{I2P} \cite{schramowski2023safe} contains 4,703 NSFW prompts sourced from real users on Lexica \cite{lexica}. The categories include hate, harassment, violence, self-harm, sex, shocking content, and illegal activities. Specifically, we categorize the prompts into hard and soft based on their level of harmfulness.
    \item \textbf{4chan Prompts} \cite{qu2023unsafe} contain 500 NSFW prompts collected from 4chan \cite{4chan}. They predominantly encompass discriminatory and derogatory statements.
    \item \textbf{NSFW200} \cite{yang2024sneakyprompt} involves 200 NSFW prompts related to sexual and bloody content.
    \item \textbf{NSFW-LAION}. We sample 1143 NSFW prompts from LAION-COCO \cite{schuhmann2022laion} to enrich datasets. These prompts mainly focus on sexual content.
\end{itemize}

\textbf{Adversarial Dataset.}
\deleted{Adversarial datasets include adversarial samples obtained through three open-source adversarial attack methods: MMA, SneakyPrompt, and Ring-A-Bell. For MMA, we utilize the 1,000 successful adversarial samples provided by the authors. For the other two methods, we generate samples with the open-source algorithms, resulting in approximately 200 samples for each.} \oldcite{yang2024mma, yang2024sneakyprompt, tsai2023ring}
\added{Adversarial datasets include adversarial prompts generated by five open-source attack methods: \textbf{MMA} \cite{yang2024mma}, \textbf{SneakyPrompt} \cite{yang2024sneakyprompt}, \textbf{Ring-A-Bell} \cite{tsai2023ring}, \textbf{JPA} \cite{jia2024jailbreaking}, and \textbf{P4D} \cite{chin2023prompting4debugging}. For MMA, we use the 1,000 successful adversarial samples released by the authors. For SneakyPrompt, Ring-A-Bell, JPA, and P4D, we generate successful attack prompts with their public implementations and evaluate them under the same protocol. Detailed settings and additional examples are deferred to the supplementary material.}

\subsubsection{Baselines}
We select eight detection methods as baselines, encompassing both prompt-based and embedding-based approaches. These include four commercial models, two open-source models, and two SOTA methods. The commercial models are OpenAI Moderation \cite{openai, markov2023holistic}, Azure AI Content Safety \cite{Azure}, AWS Comprehend \cite{AWS}, and Aliyun Text Moderation \cite{Ali}. These systems primarily rely on large transformer-based architectures to identify potential toxic content in prompts. The open-source moderators, NSFW-text-classifier \cite{NSFWtext} and Detoxify \cite{Detoxify}, leverage lightweight models to detect NSFW text content, offering faster inference speeds. Latent Guard \cite{liu2024latent} and GuardT2I \cite{yang2024guardt2i} are embedding-based methods that represent the current SOTA in detecting NSFW prompts for T2I models.


\begin{table*}[t]
    \centering
    \caption{The overall evaluation of AEIOU.}
    \label{tab:overall}
    \resizebox{\textwidth}{!}{%
    \begin{tabular}{lcccccccc}
        \toprule
        Detector & TPR & FPR & ACC & F1 Score & AUROC & AUPRC & TPR@FPR 1\% & Time/Query(ms) \\
        \midrule
        OpenAI Moderation \cite{openai} & 0.2976 & \textbf{0.0010} & 0.8220 & 0.4578 & 0.8616 & 0.7960 & 0.4974 & 1288.43 \\
        Azure AI Content Safety \cite{Azure} & 0.4761 & 0.0118 & 0.8590 & 0.6302 & 0.7331 & 0.7708 & 0.4313 & 922.67 \\
        AWS Comprehend \cite{AWS} & 0.4702 & 0.0730 & 0.8118 & 0.5576 & 0.7143 & 0.6244 & 0.2980 & 286.42 \\
        Aliyun Text Moderation \cite{Ali} & 0.1736 & 0.0023 & 0.7897 & 0.2941 & 0.5856 & 0.6720 & 0.1799 & 99.16 \\
        NSFW-text-classifier \cite{NSFWtext} & 0.7325 & 0.3534 & 0.6699 & 0.5466 & 0.7627 & 0.6823 & 0.2922 & 9.14 \\
        Detoxify \cite{Detoxify} & 0.5432 & 0.1778 & 0.7465 & 0.5379 & 0.7226 & 0.6455 & 0.3340 & 24.82 \\
        Latent Guard \cite{liu2024latent} & 0.5021 & 0.1403 & 0.7625 & 0.5346 & 0.7579 & 0.5995 & 0.1690 & 167.90 \\
        GuardT2I \cite{yang2024guardt2i} & 0.7102 & 0.0779 & 0.8686 & 0.7318 & 0.9160 & 0.8207 & 0.3492 & 352.3 \\
        \vspace{-1em} \\
        \hline
        \vspace{-0.8em} \\
        AEIOU$_{\text{CLIP-L}}$ & \textbf{0.9833} & 0.0085 & \textbf{0.9895} & \textbf{0.9792} & \textbf{0.9990} & \textbf{0.9977} & \textbf{0.9842} & \textbf{0.64} \\
        AEIOU$_{\text{CLIP-G}}$ & 0.9747 & 0.0082 & 0.9875 & 0.9751 & \textbf{0.9990} & 0.9974 & 0.9799 & 1.84 \\
        AEIOU$_{\text{T5}}$ & 0.9726 & 0.0102 & 0.9853 & 0.9708 & 0.9984 & 0.9957 & 0.9701 & 6.71 \\
        AEIOU$_{\text{ua}}$ & 0.9829 & 0.0087 & 0.9873 & 0.9785 & 0.9982 & 0.9971 & 0.9824 & \textbf{0.64} \\
        AEIOU$_{\text{multi}}$ & 0.9763 & 0.0066 & 0.9890 & 0.9783 & 0.9988 & 0.9973 & 0.9817 & 0.93 \\
        \bottomrule
    \end{tabular}
    }
    \begin{minipage}{\linewidth}
    \vspace{0.5em}
    \small Note: AEIOU$_{\text{CLIP-L}}$, AEIOU$_{\text{CLIP-G}}$ and AEIOU$_{\text{T5}}$ are methods deployed on three different text encoders. AEIOU$_{\text{ua}}$ is trained without any adversarial prompts, while AEIOU$_{\text{multi}}$ is trained across multiple categories and integrates the results. Both of them are deployed on CLIP-L.
    \end{minipage}
\end{table*}

\subsubsection{Metrics}

We evaluate our model using a standard suite of binary classification metrics, including the True Positive Rate (TPR), the False Positive Rate (FPR), overall Accuracy, and the F1 Score \cite{powers2011evaluation}. For a threshold-agnostic analysis, we report the Area Under the ROC Curve (AUROC) for overall discriminative power and the Area Under the Precision-Recall Curve (AUPRC), which is more informative for skewed class distributions \cite{fawcett2006introduction}. To assess performance under strict operational constraints, we also measure TPR at a low FPR of 1\%. The decision threshold for these point-based metrics is determined by optimizing the F1 score on the training set. \deleted{Finally, we assess the runtime efficiency of our method, AEIOU, by reporting its average inference time per query.}
\added{Finally, we report the average inference time per query in the main paper. A more complete comparison of latency, throughput, and peak GPU memory is provided in the supplementary material.}

\subsubsection{Implementation Details}
We deploy AEIOU on three commonly used text encoders: CLIP-ViT-L (\textbf{CLIP-L}) \cite{clipl}, CLIP-ViT-bigG (\textbf{CLIP-G}) \cite{clipg}, and T5-v1.1-XXL (\textbf{T5}) \cite{t5}. CLIP-L is the most widely used text encoder and serves as the foundation for most safety-related research. CLIP-G and T5 are larger models; many recent models \cite{flux, esser2024scaling} employ them as text encoders.

We design two variant detectors to evaluate our method's effectiveness in different scenarios. First, when the defender is unaware of the attack method, we employ a model trained solely on clean data and regular NSFW data to test AEIOU's generalization capability. Second, when the defender has sufficient data labeled with specific categories, they can train on multiple specific concepts to enhance the model's specificity. We deploy the CLIP-L-based detector in both scenarios. \added{Unless otherwise specified, the local experiments are conducted on an NVIDIA A800 (80GB) GPU and an Intel Xeon Platinum 8470 CPU with batch size 64 in FP32 precision.}

\begin{table*}[t]
    \centering
    \caption{The accuracy of each dataset.}
    \label{tab:sets}
    \resizebox{\textwidth}{!}{%
    \begin{tabular}{lcccccccccc}
        \toprule
        Detector & I2P-Soft & I2P-Hard & 4chan & NSFW200 & NSFW-laion & MMA & SneakyPrompt & Ring-A-Bell & \added{JPA} & \added{P4D} \\
        \midrule
        OpenAI Moderation \cite{openai} & 0.0244 & 0.0600 & 0.8200 & 0.6800 & 0.3147 & 0.7030 & 0.6311 & 0.6117 & \added{0.6438} & \added{0.6925} \\
        Azure AI Content Safety \cite{Azure} & 0.1184 & 0.2162 & 0.9920 & 0.8300 & 0.8287 & 0.8176 & 0.8058 & 0.9126 & \added{0.7326} & \added{0.8418} \\
        AWS Comprehend \cite{AWS} & 0.1462 & 0.2303 & \textbf{1.0000} & 0.7900 & 0.5909 & 0.8977 & 0.7699 & 0.8252 & \added{0.7217} & \added{0.8081} \\
        Aliyun Text Moderation \cite{Ali} & 0.0627 & 0.0415 & 0.9200 & 0.1700 & 0.3392 & 0.1502 & 0 & 0.5437 & \added{0.3742} & \added{0.6210} \\
        NSFW-text-classifier \cite{NSFWtext} & 0.4930 & 0.5841 & \textbf{1.0000} & 0.9700 & 0.7972 & 0.9644 & 0.9029 & 0.9806 & \added{0.8244} & \added{0.9167} \\
        Detoxify \cite{Detoxify} & 0.2180 & 0.3384 & \textbf{1.0000} & 0.8300 & 0.4248 & 0.9377 & 0.7282 & 0.8544 & \added{0.7650} & \added{0.8132} \\
        Latent Guard \cite{liu2024latent} & 0.2451 & 0.3581 & 0.9720 & 0.3600 & 0.5455 & 0.7753 & 0.2427 & 0.5922 & \added{0.4571} & \added{0.6355} \\
        GuardT2I \cite{yang2024guardt2i} & 0.5926 & 0.6157 & 0.8640 & 0.8100 & 0.7902 & 0.8365 & 0.8835 & \textbf{1.0000} & \added{0.8043} & \added{0.8567} \\
        \vspace{-1em} \\
        \hline
        \vspace{-0.8em} \\
        AEIOU$_{\text{CLIP-L}}$ & 0.9652 & 0.9858 & 0.9800 & 0.9800 & 0.9983 & \textbf{0.9989} & \textbf{0.9903} & \textbf{1.0000} & \textbf{\added{0.9892}} & \added{0.9911} \\
        AEIOU$_{\text{CLIP-G}}$ & 0.9673 & 0.9825 & 0.9960 & \textbf{0.9900} & 0.9930 & 0.9978 & \textbf{0.9903} & \textbf{1.0000} & \added{0.9826} & \textbf{\added{0.9938}} \\
        AEIOU$_{\text{T5}}$ & 0.9519 & 0.9738 & 0.9680 & 0.9600 & 0.9843 & 0.9956 & 0.9806 & \textbf{1.0000} & \added{0.9826} & \added{0.9874} \\
        AEIOU$_{\text{ua}}$ & \textbf{0.9680} & \textbf{0.9891} & 0.9800 & 0.9700 & 0.9979 & 0.9911 & 0.9806 & \textbf{1.0000} & \added{0.9852} & \textbf{\added{0.9938}} \\
        AEIOU$_{\text{multi}}$ & 0.9547 & 0.9705 & 0.9880 & 0.9800 & \textbf{1.0000} & 0.9933 & \textbf{0.9903} & \textbf{1.0000} & \added{0.9852} & \added{0.9911} \\
        \bottomrule
    \end{tabular}
    }
\end{table*}

\subsection{Overall Evaluation}
We first conduct an overall evaluation of the effectiveness of AEIOU. We deploy it across three different text encoders and trained models using three distinct settings. It is then compared against eight baselines on one benign and eight NSFW datasets. 
We randomly selected 2,000 prompts from a total of over 35,000 prompts to serve as the training set and used the remaining prompts as the test set.

Table \ref{tab:overall} presents the overall evaluation results across all datasets, with the best performance for each metric highlighted in bold. As shown in the table, AEIOU consistently outperforms previous classification approaches across almost all metrics. It demonstrates high detection accuracy across the CLIP-L, CLIP-G, and T5 models, proving its applicability to various text encoders.

Table \ref{tab:sets} \deleted{shows accuracy across datasets, with the first six columns covering regular NSFW datasets and the last three focusing on adversarial datasets.} \added{shows accuracy across datasets, with the first five columns covering regular NSFW datasets and the remaining columns focusing on adversarial prompt datasets from multiple attack sources.} AEIOU maintains high accuracy across all datasets, whereas the performance of other methods is inconsistent. Most methods perform poorly on the I2P dataset. This is likely because they need to detect text across various applications, making it hard to handle prompts used in T2I models specifically. On other regular NSFW datasets, most classifiers achieve relatively high accuracy but still lag behind AEIOU. Notably, on the 4chan dataset, three classifiers achieve higher accuracy than AEIOU. However, the difference is very minimal. Finally, regarding adversarial prompts, AEIOU significantly outperforms all other methods. Even AEIOU$_{\text{ua}}$, which is trained without adversarial datasets, still achieves remarkably high accuracy. \added{The newly added JPA and P4D columns further broaden the adversarial evaluation beyond the original benchmark set and test whether the detector remains reliable across more diverse prompt-generation mechanisms.}

Additionally, we assess each method's efficiency by measuring the average time per query, presented in the last column of Table \ref{tab:overall}. AEIOU requires significantly less time than other methods, primarily because they often utilize large models for detection. In contrast, AEIOU only incorporates multiple matrix operations during the text encoder's inference process. On the smallest CLIP-L model, AEIOU's efficiency improves at least tenfold compared to other models. Even on larger models like CLIP-G and T5, AEIOU's efficiency surpasses that of all other models. \added{A more complete runtime analysis, including throughput and peak GPU memory, is provided in the supplementary material.}

To comprehensively illustrate accuracy across a range of thresholds, we plot the ROC curves for AEIOU and the baseline methods, as shown in Figure \ref{fig: roc}. ROC curve for AEIOU is consistently positioned above other methods. This demonstrates that AEIOU maintains superior performance compared to the baselines across a wide spectrum of classification thresholds.

\subsection{Generalization to Unknown Attacks}
This section examines AEIOU's ability to defend against unknown adversarial attacks. In Tables \ref{tab:overall} and \ref{tab:sets}, we present the performance of AEIOU trained solely on benign and regular NSFW datasets, denoted as AEIOU$_{\text{ua}}$. Despite never encountering adversarial prompts, the experimental results indicate that AEIOU$_{\text{ua}}$ can still effectively identify adversarial NSFW prompts, with accuracy only slightly lower than the standard AEIOU. We attribute this effectiveness to AEIOU's focus on the semantic information embedded in the hidden states. Although adversarial and regular NSFW prompts may appear different to the human eye, their semantic information is similar, allowing AEIOU$_{\text{ua}}$ to recognize them accurately.

\begin{table*}[ht]
    \centering
    \caption{The evaluation of different categories.}
    \label{tab:concepts}
    \resizebox{0.8\textwidth}{!}{%
    \begin{tabular}{lccccccc}
        \toprule
        Detector & Sexual & Hate & Self-Harm & Violence & Shocking & Harassment & Illegal \\
        \midrule
        AEIOU & 0.9959 & \textbf{0.9863} & 0.9763 & 0.9921 & 0.9790 & \textbf{0.9733} & 0.9340 \\
        AEIOU$_{\text{multi}}$ & \textbf{0.9991 }& 0.9795 & \textbf{0.9950} & \textbf{0.9974} & \textbf{0.9848} & 0.9697 & \textbf{0.9725} \\
        \bottomrule
    \end{tabular}
    }
\end{table*}

\begin{table*}[t]
    \centering
    \caption{The impact of training data size.}
    \label{tab:data_size}
    \resizebox{0.8\textwidth}{!}{%
    \begin{tabular}{lccccccc}
        \toprule
        Training Data Size & TPR   & FPR   & ACC   & F1 Score & AUROC & AUPRC & TPR@FPR 1\% \\
        \midrule
        10   & 0.9171 & 0.0278 & 0.9583 & 0.9174 & 0.9875 & 0.9734 & 0.8495 \\
        50   & 0.9356 & 0.0164 & 0.9714 & 0.9430 & 0.9940 & 0.9861 & 0.9157 \\
        100  & 0.9605 & 0.0180 & 0.9766 & 0.9538 & 0.9962 & 0.9906 & 0.9351 \\
        500  & 0.9824 & \textbf{0.0091} & \textbf{0.9888} & \textbf{0.9778} & 0.9981 & 0.9974 & 0.9829 \\
        1000 & \textbf{0.9833} & 0.0096 & 0.9887 & 0.9776 & \textbf{0.9990} & \textbf{0.9975} & \textbf{0.9833} \\
        \bottomrule
    \end{tabular}
    }
\end{table*}

\begin{figure*}[t]
    \centering
    \includegraphics[width=1.0\linewidth]{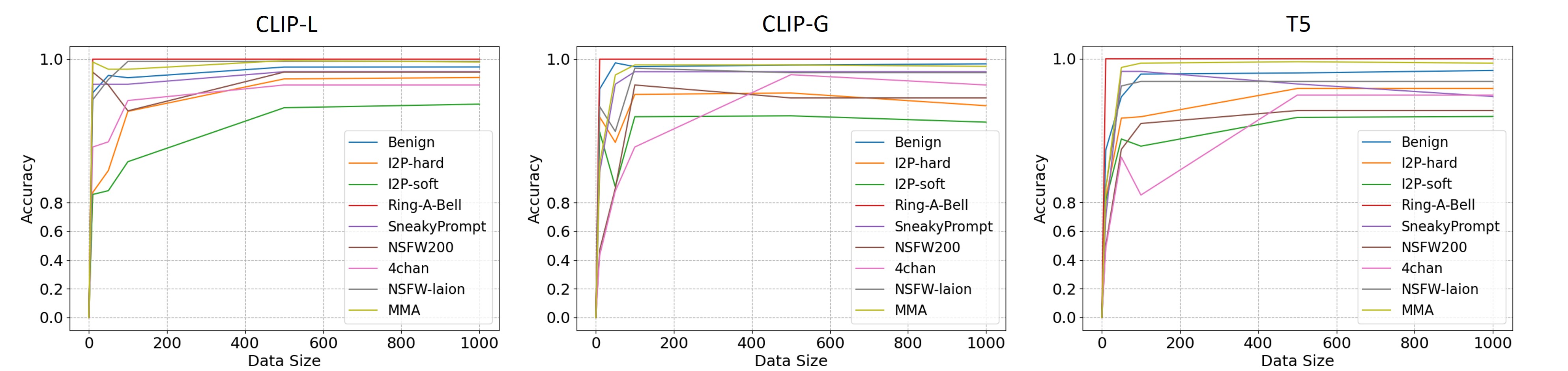}
    \caption{The impact of training data size.}
    \label{fig: data_size}
\end{figure*}

\subsection{Multi-Categories Classifier}
NSFW serves as an overarching descriptor for harmful prompts, and it can be decomposed into more specific categories. In this section, we follow the I2P dataset to classify NSFW prompts into seven particular categories: sexual, hate, self-harm, violence, shocking, harassment, and illegal. Adhering to the methodology described in Section \ref{sec:detect}, we identify features representing these concepts and derive the multi-categories AEIOU by integrating the NSFW scores of each category. Tables \ref{tab:overall} and \ref{tab:sets} compare the overall performance of AEIOU$_{\text{multi}}$ with the standard AEIOU. Although the detailed AEIOU$_{\text{multi}}$ exhibits slightly worse overall performance, the difference is minimal. Furthermore, Table \ref{tab:concepts} compares their accuracies within each category. AEIOU$_{\text{multi}}$ generally achieves higher accuracy in most categories, but in some, it underperforms compared to the standard AEIOU. 

We attribute the lack of superiority in AEIOU$_{\text{multi}}$ to two main reasons. First, prompts from different categories often share overlapping features. Common sensitive words like ``f**k" appear across multiple categories, which limits AEIOU$_{\text{multi}}$'s ability to capture shared features when trained individually on each category. As a result, it demonstrates higher accuracy in more distinct categories like self-harm but lower accuracy in more ambiguous categories such as hate and harassment. Secondly, discrepancies in data quality exist among different categories. In our datasets, sexual prompts have broad coverage and the highest quality, while the quality of other categories' prompts varies significantly. This leads to suboptimal performance of AEIOU\(_{\text{multi}}\) on some categories. To improve AEIOU\(_{\text{multi}}\)'s performance, we need to train it using higher-quality datasets.

\subsection{The Impact of Training Data Size}
In this section, we discuss the impact of training data size on the performance of AEIOU. We set the training data size for AEIOU to 10, 50, 100, 500 and 1000. Half of the training data is randomly selected benign data, while the other half is randomly selected NSFW data. Table \ref{tab:data_size} presents the experimental results of AEIOU$_{\text{CLIP-L}}$. The experiments demonstrate that AEIOU maintains high accuracy even with only 10 training samples. When the sample size reaches 500, its performance is comparable to AEIOU trained with a full dataset. As the number of training samples increases further, there is no significant improvement in performance, indicating that improving the quality and coverage of training samples is a better strategy than simply increasing the quantity. Figure \ref{fig: data_size} provides additional insights with different training data sizes across various datasets, which aligns with the results in Table \ref{tab:data_size}. These experiments confirm AEIOU's exceptional performance in few-shot scenarios.

\begin{figure*}[ht]
    \centering
    \includegraphics[width=1.0\linewidth]{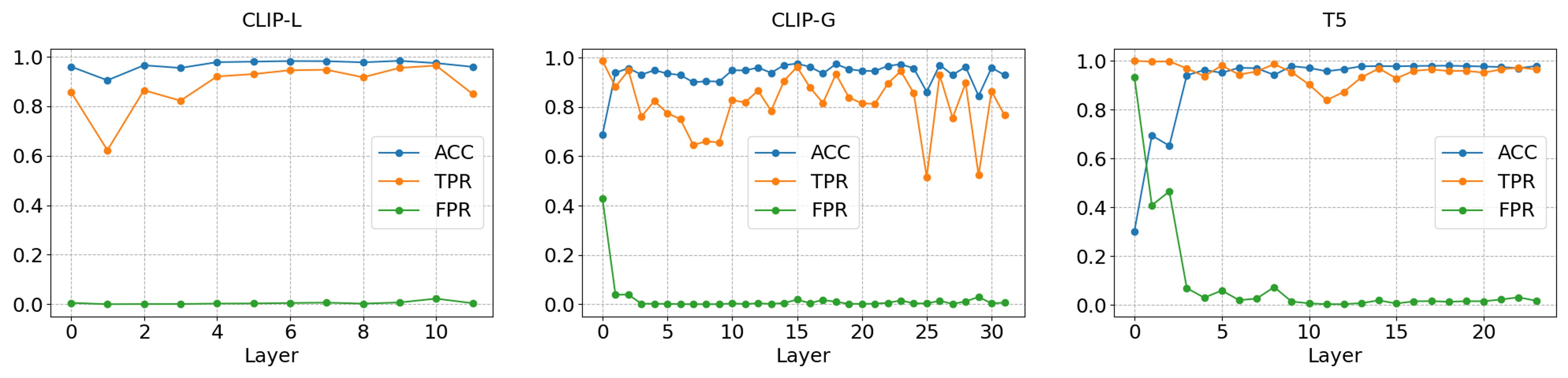}
    \caption{Effectiveness of each layer.}
    \label{fig: ablation}
\end{figure*}

\begin{table*}[ht]
    \centering
    \caption{The evaluation of adaptive attack.}
    \label{tab:adaptive}
    \resizebox{0.8\textwidth}{!}{%
    \begin{tabular}{lccccccccc}
        \toprule
        \multirow{2}{*}{Defender} & \multicolumn{3}{c}{SneakyPrompt} & \multicolumn{3}{c}{MMA} \\
        \cmidrule(lr){2-4} \cmidrule(lr){5-7}
         & ASR-H(\%) & ASR-M(\%) & CLIP Score & ASR-H(\%) & ASR-M(\%) & CLIP Score \\
        \midrule
        Bare & 46 & 48 & 0.9597 & 75 & 79 & 0.9315 \\
        ESD & 5 & 7 & 0.9613 & 10 & 13 & 0.9315 \\
        AEIOU & 0 & 0 & / & 28 & 31 & 0.8850 \\
        AEIOU+ESD & 0 & 0 & / & 0 & 0 & / \\ 
        AEIOU$_{\text{DA}}$ & 0 & 0 & / & 6 & 7 & 0.8482 \\
        AEIOU$_{\text{DA}}$+ESD & 0 & 0 & / & 0 & 0 & / \\
        \bottomrule
    \end{tabular}
    }
\end{table*}

\subsection{The Ablation Study}
In our approach, we utilize NSFW features from all layers and all attention heads for detection. In this section, we will discuss the impact of using NSFW features from only a single layer of text encoder. We conduct experiments on three text encoders. Figure \ref{fig: ablation} presents the accuracy, TPR, and FPR when using each layer for detection. 

In all three text encoders, even when using attention heads from a single layer, many layers still achieve high accuracy. For CLIP-L and CLIP-G, the middle layers tend to have higher accuracy, while the early and final layers show lower accuracy. Conversely, in the T5 model, the later layers exhibit higher accuracy. This highlights the distinct characteristics of the two types of text encoders.

Although using attention heads from a single layer can achieve high accuracy, we recommend using the original AEIOU method for the highest precision in detection.

\section{Adaptive Attack}
In this section, we evaluate the robustness of AEIOU against adaptive attacks. To target our model, we design adaptive attacks based on SneakyPrompt \cite{yang2024sneakyprompt} and MMA \cite{yang2024mma}. They are applicable to all T2I models and can effectively bypass both internal and external safeguards. Considering their different applicable scenarios, we employ SneakyPrompt for black-box adaptive attack and MMA for white-box adaptive attack.

We utilize Stable Diffusion v1.4 \cite{sd1.4} as the generative model, as it is the model most vulnerable to attacks. We evaluate the attack performance on the bare model, the ESD model \cite{gandikota2023erasing}, and the model employing AEIOU. The bare model only detects whether sensitive tokens are present in the prompt, while ESD fine-tunes the model to make it difficult to generate NSFW images. We assess the attacks using two metrics: attack success rate (ASR) and CLIP Score \cite{hessel2021clipscore}. 
The attack success rate evaluates whether adversarial attacks can successfully generate NSFW images. The CLIP Score assesses the semantic similarity between adversarial and target prompts. The lower the CLIP score, the further the adversarial prompt deviates from the semantic meaning of the target prompt.

To ensure result reliability, both model-based and manual evaluations are conducted for attack success rate. A SOTA model \cite{qu2023unsafe} is used for automated classification, while human evaluations are performed independently by three individuals, with the majority opinion determining the final assessment.


\subsection{Black-Box Adaptive Attack}
In the black-box scenario, we assume the attacker has no knowledge of the model's details but can choose prompts and query the model to obtain output. We integrate AEIOU into the text encoder of Stable Diffusion. When a potential NSFW prompt is detected, the model will refuse to generate the image. We select 100 prompts with clear NSFW semantics from NSFW200 as target prompts and use SneakyPrompt to attack the model. For all prompts successfully generated by SneakyPrompt, we then input them into the diffusion model to test whether they can generate NSFW images. Each image classified as NSFW is counted as a single success. Experimental results are shown in Table \ref{tab:adaptive}. ASR-H and ASR-M represent the ASR evaluated by human and classification model, respectively.

When attacking the bare model, SneakyPrompt achieves a high attack success rate. ESD significantly reduces the attack success rate, yet it cannot entirely prevent the generation of NSFW images. However, after incorporating AEIOU, SneakyPrompt is entirely thwarted and unable to generate any NSFW images. This is because SneakyPrompt only replaces a few tokens in the prompt, which does not effectively neutralize its overall meaning.

\subsection{White-Box Adaptive Attack}
In a white-box scenario, we assume attackers can only generate images through queries. However, they possess a local copy of the text encoder identical to the target model and are aware of the AEIOU defense strategy. The attacker can target AEIOU by modifying the loss function to conduct a specific attack. The MMA attack's loss aims to make the conditional embeddings of the adversarial prompt and the target prompt as similar as possible. To effectively attack AEIOU, we incorporate $Score(p)$ as $L_{\text{AEIOU}}$ into the original loss function. This transforms the objective of the loss function to minimize the NSFW Score while ensuring that the semantics of the adversarial prompt closely align with the target prompt.
\begin{equation}
    L = L_{\text{MMA}} + \lambda \times L_{\text{AEIOU}}
\end{equation}
Where $\lambda$ is a weighting factor that balances between the two components. Based on this foundation, we implement a target truncation strategy. Specifically, once $L_{\text{AEIOU}}$ exceeds the threshold by a small margin, we stop optimizing it and shift our primary focus to optimizing $L_{\text{MMA}}$. This enables adversarial prompts to approximate the semantics of the target prompt as closely as possible while avoiding detection. Consequently, the final loss is formulated as:

\begin{equation}
L = L_{\text{MMA}} + \lambda \times \max(L_{\text{AEIOU}}, \tau - \epsilon),
\end{equation}

Where $\tau$ is the threshold and $\epsilon$ is the margin. When updating the best prompt, we ensure that the $L_{\text{AEIOU}}$ surpasses the threshold. We conduct the attack using the default settings of MMA and evaluate it on 100 target prompts. The experimental results are presented in Table \ref{tab:adaptive}.

As a white-box attack, MMA demonstrates stronger capabilities than SneakyPrompt on the bare model. However, when it targets AEIOU, AEIOU exhibits robust performance, significantly reducing the attack success rate. Meanwhile, the average CLIP score of successfully generated adversarial prompts also decreases, indicating that MMA's optimization of these prompts is not as successful as in the bare model. Moreover, although AEIOU itself does not increase the memory usage of the text encoder, the adaptive attack against it must optimize hidden states across all attention heads, significantly increasing the memory requirements. While standard MMA operates on less than 10GB of memory, adaptive MMA demands nearly 50GB, restricting its execution to commercial-grade GPUs. This makes adaptive attacks on AEIOU more challenging.

We can also combine AEIOU with other defense methods. For instance, by integrating ESD, we can reduce the success rate of MMA attacks to zero. By implementing a comprehensive defense strategy that addresses other parts of the T2I model, we can enhance the overall defensive performance and make adaptive attacks more challenging.

To further enhance AEIOU's resilience against adaptive attacks, we incorporate adversarial prompts that successfully breach AEIOU's defenses into the training set for additional training. The following section will provide a detailed discussion of this process.

\subsection{Red-Teaming Data Augmentation}
We conduct red team testing using a white-box adaptive attack and collect adversarial prompts that successfully bypass AEIOU and generate NSFW images. These prompts are added to AEIOU's training dataset for data augmentation, aiming to enhance AEIOU's ability to resist corresponding adaptive attacks. When training, we assign them greater weight to ensure they significantly influence the model even if the sample size is limited.

Table \ref{tab:adaptive} presents the performance of the data-augmented AEIOU. We include 25 adaptive adversarial prompts in the training set with a weight of 50. This results in AEIOU$_{\text{DA}}$, demonstrating significantly improved defense against adaptive attacks, reducing the success rate to just 7\% for MMA. Additionally, the CLIP score experiences a further decline. This underscores the effectiveness of further optimizing AEIOU with data augmentation.

In the practical application of AEIOU, we can also collect adversarial prompts that successfully bypass defenses through manual screening. These prompts can then be added to the training dataset with appropriate weighting, allowing for continuous updates of the defense model.

\section{Interpretation Experiments}
In this section, we validate AEIOU's interpretability. This not only makes AEIOU's classification more transparent and trustworthy but also aids users in further understanding the semantics of prompts. We assess the reliability of interpretations across both text and image modalities.

\subsection{Text-Based Interpretation}
\begin{figure}[t]
    \centering
    \includegraphics[width=0.8\linewidth]{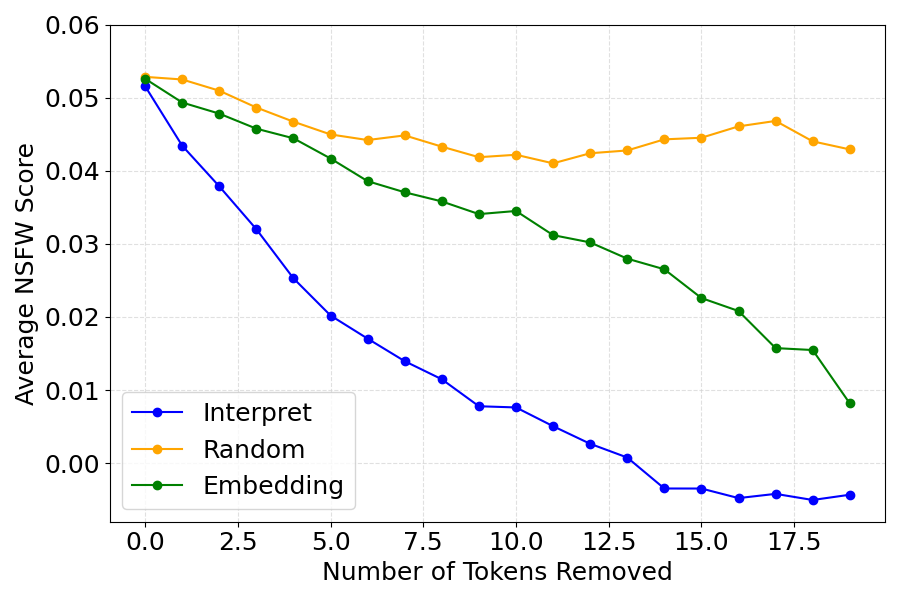}
    \caption{Text-based interpretation.}
    \label{fig: text_int}
\end{figure}

In the text modality, our interpretation aids users in understanding the semantics of a prompt by identifying tokens containing NSFW semantics. After obtaining the interpretation result \( E(p)_i \) for each token in the prompt \( p \), we sequentially remove the corresponding tokens from \( p \) in descending order of \( E(p)_i \) and observe the changes in the NSFW score. We compare our interpretation method against two baselines. The first, a random-based method, removes tokens from the prompt at random. The second, an embedding-based method, uses the CLIP score to evaluate the semantic similarity of each token to the term "NSFW" and removes tokens in descending order of this similarity. The experimental results are shown in Figure \ref{fig: text_int}. Compared to both baselines, removing tokens based on our interpretation method more rapidly reduces the NSFW semantic content of the prompt. This demonstrates the effectiveness of AEIOU in both explaining NSFW prompts and locating the specific tokens responsible for the NSFW content.


Additionally, we conduct further experiments on image generation using the 100 prompts mentioned above. We compare the images generated from the original prompt with those generated from the prompt after removing the NSFW token. For original prompts, 87 out of the 100 generated images are NSFW. After removing NSFW tokens, only 5 prompts result in NSFW images. This further demonstrates the effectiveness of our interpretation method.

\subsection{Image-Based Interpretation}
Directly removing tokens with NSFW semantics can prevent the generation of NSFW images. However, it significantly disrupts the original intent of the prompt. To address this drawback, we manipulate the embeddings of each token in the image-based interpretation process, gradually eliminating NSFW semantics and observing changes in the generated images. We use the same 100 prompts as in the previous section. We evaluate the degree to which the generated images contain NSFW semantics and how closely they align with the original prompt semantics.

\begin{figure}[t]
    \centering
    \includegraphics[width=0.8\linewidth]{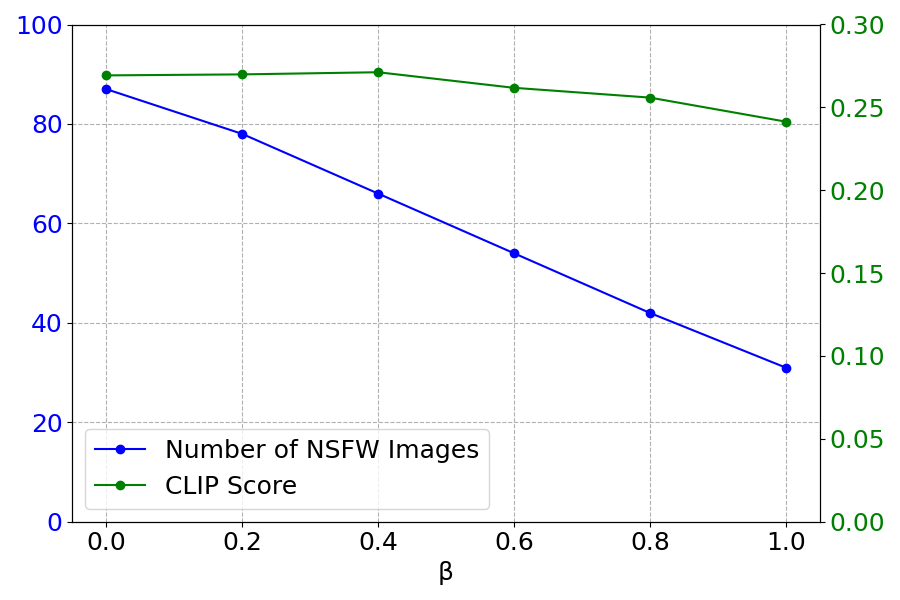}
    \caption{Image-based interpretation.}
    \label{fig: image_int}
\end{figure}

Figure \ref{fig: image_int} illustrates the variation in the number of NSFW images generated and the semantic similarity between the images and prompts as the parameter $\beta$ changes. The semantic similarity is evaluated using the CLIP Score. As $\beta$ increases, the number of NSFW images decreases progressively. Although the CLIP Score also shows a decreasing trend, the overall deviation from the original image's CLIP Score remains minimal. This indicates that the NSFW semantics of the prompt are effectively mitigated while preserving other semantic information as much as possible.

However, because this method directly modifies semantics within the hidden states, the resulting conditional embeddings deviate from the normal distribution, often leading to lower-quality images. Therefore, this approach is supposed to aid in understanding the representation of NSFW semantics within the text encoder. It cannot be directly applied to erase specific concepts from images.

\section{Discussion}
\textbf{Practicality.} Previous research \cite{qi2024ai} has identified key qualities for a safe and secure generative model, including integrity, robustness, alignment, and interpretability. Building on this foundation, we propose that a practical defense framework should adhere to the following principles: First, the defense must be integrated, offering protection against both conventional threats and adversarial attacks. Second, the defense should remain robust amidst changing external conditions, addressing issues such as distribution shifts and adaptive attacks. Third, the model with the defensive mechanism should align with the original model, preserving its effectiveness and efficiency. Lastly, the defense should be interpretable, enabling the timely identification of anomalous behaviors to prevent potential hazards.

AEIOU excels in these four areas compared to prior methods. Firstly, AEIOU maintains an accuracy rate exceeding 95\% across various models, significantly outperforming earlier approaches. Secondly, it requires only a small number of samples for data augmentation, simplifying updates. Thirdly, its classification process is transparent, and it provides interpretative tools to help users understand target prompts. Lastly, AEIOU does not impact image generation quality and incurs only negligible computational overhead. These all demonstrate its practicality.

\hspace{-1em}\textbf{Limitations and Future Work.} AEIOU extracts NSFW features from hidden states, showcasing the potential for extracting specific concepts from the text encoder. However, this study does not explore the extraction of other concepts, which warrants further investigation.

Due to its low computational cost, AEIOU utilizes all attention heads for detection. However, to better understand the unique characteristics of text encoders like CLIP, it is essential to analyze the semantic focus of individual attention heads. Exploring the roles and properties of each attention head remains a valuable direction for future research.

Despite AEIOU demonstrating robust capabilities, no defensive method is foolproof. Therefore, in practical applications, it should be combined with other techniques, such as image moderation. Given that CLIP is a multimodal model, there is potential for adapting AEIOU to the image domain to achieve more powerful post-generation defenses.


\section{Conclusion}
In this paper, we propose a unified defense framework against NSFW prompts in T2I models named AEIOU. The AEIOU framework, being adaptable, efficient, interpretable, optimizable, and unified, has demonstrated superior performance that far exceeds previous defense methods. In addition to detection, we provide interpretability methods to help understand the semantics of NSFW prompts and the generation process of NSFW images. Experimental results show that AEIOU exhibits strong capabilities in defending against both normal attacks and adaptive attacks. 

\bibliographystyle{IEEEtran}
\bibliography{reference}

\newpage

\begin{IEEEbiographynophoto}{Yiming Wang}
Yiming Wang received the BS from the College of Computer Science and Technology, Zhejiang University, in 2023. He is currently working toward the PhD degree in the College of Computer Science and Technology with Zhejiang University. His research interests include AI safety and security, especially AIGC safety.
\end{IEEEbiographynophoto}

\begin{IEEEbiographynophoto}{Jiahao Chen}
Jiahao Chen received the BS from the Faculty of Electrical Engineering and Computer Science, Ningbo University, in 2023. He is currently working toward the PhD degree in the College of Computer Science and Technology with Zhejiang University. His research interests include trustworthy generative AI in real-world applications, especially AI security, AI copyright protection, and privacy-preserving AI.
\end{IEEEbiographynophoto}

\begin{IEEEbiographynophoto}{Qingming Li}
Qingming Li received the B.S. degree from Xidian University, Xi'an, China, in 2017, and the Ph. D degree from Tsinghua University, Beijing, China, in 2022. She was a Research Assistant with the Research Institute of Artificial Intelligence, Zhejiang Lab, Hangzhou, China from July 2022 to Mar. 2024. Since Apr. 2024, she has been a Postdoc with the College of Computer Science and Technology, Zhejiang University, Hangzhou, China. Her research interests include deep learning security, large language models, and embodied agents. 
\end{IEEEbiographynophoto}

\begin{IEEEbiographynophoto}{Tong Zhang}
Tong Zhang is a Ph.D. student in the College of Computer Science and Technology, Zhejiang University. He received his bachelor's degree from Shandong Agricultural University in 2021, and his master's degree from the School of Software, Shandong University in 2024. His research interests include adversarial attacks,  copyright protection,post-training of multimodal large language models, and video grounding.
\end{IEEEbiographynophoto}

\begin{IEEEbiographynophoto}{Rui Zeng}
Rui Zeng received the bachelor's degree from Huazhong University of Science and Technology. He is currently working toward the PhD degree with the College of Computer Science and Technology, Zhejiang University. His current research interests include AI security and software security.
\end{IEEEbiographynophoto}

\begin{IEEEbiographynophoto}{Xing Yang}
Xing Yang comes from Dujiangyan City, Chengdu, China. He received the bachelor’s, master’s and doctor’s degrees from Hefei Electronic Engineering Institute in 2006, 2009, and 2012 respectively. At present, he is a researcher with the State Key Laboratory of Pulsed Power Laser Technology, National University of Defense Technology and mainly interested in the research fields of optoelectronic engineering, artificial intelligence, and cyberspace security.
\end{IEEEbiographynophoto}

\begin{IEEEbiographynophoto}{Shouling Ji}
Shouling Ji (Member, IEEE) received the PhD degree in electrical and computer engineering from the Georgia Institute of Technology and the PhD degree in computer science from Georgia State University. He is a Qiushi distinguished professor with the College of Computer Science and Technology, Zhejiang University. His current research interests include Data-driven Security and Privacy, AI Security and Software and System Security. He is a member of ACM and a senior member of CCF. He was a research intern with the IBM T. J. Watson Research Center. He is the recipient of the 2012 Chinese Government Award for Outstanding Self-Financed Students Abroad and ten Best/Outstanding Paper Awards, including ACM CCS 2021.
\end{IEEEbiographynophoto}

\end{document}